# AlN/Si interface engineering to mitigate RF losses in MOCVD grown GaN-on-Si substrates

Pieter Cardinael[1,#], Sachin Yadav[2], Herwig Hahn[3], Ming Zhao[2], Sourish Banerjee[2], Babak Kazemi Esfeh[2], Christof Mauder[3], Barry O'Sullivan[2], Uthayasankaran Peralagu[2], Anurag Vohra[2], Robert Langer[2], Nadine Collaert[2], Bertrand Parvais[2,4], Jean-Pierre Raskin[1]

[1]Institute for Information and Communication Technologies, Electronics and Applied Mathematics, Université catholique de Louvain, Place du Levant 3, 1348 Louvain-la-Neuve, Belgium
[2]IMEC, Kapeldreef 75, 3001 Leuven, Belgium
[3]AIXTRON SE, Dornkaulstr. 2, 52134 Herzogenrath, Germany
[4]Department of Electronics and Informatics, Vrije Universiteit Brussels, 1050 Ixelles, Belgium
[#]Corresponding author: pieter.cardinael@uclouvain.be

**Abstract.**
Fabrication of low-RF loss GaN-on-Si HEMT stacks is critical to enable competitive front-end-modules for 5G and 6G applications. The main contribution to RF losses is the interface between the III-N layer and the HR Si wafer, more specifically the AlN/Si interface. At this interface, a parasitic surface conduction layer exists in Si, which decreases the substrate effective resistivity sensed by overlying circuitry below the nominal Si resistivity. However, a clear understanding of this interface with control of the parasitic channel is lacking. In this letter, a detailed physical and electrical description of MOCVD-grown AlN/Si structures is presented. The presence of a $SiC_xN_y$ interfacial layer is revealed and its importance for RF losses is shown. Through C-V and I-V characterisation, an increase in the C concentration of this interfacial layer is linked to the formation of negative charge at the AlN/Si interface, which counteracts the positive charge present in the 0-predose limit. The variation of TMAl predose is shown to allow precise tuning of the C composition and, consequently, the resulting interface charge. Notably, a linear relationship between predose and net interface charge is observed and confirmed by the fabrication of an AlN/Si sample with close to zero net charge. In addition, a higher $D_{it}$ ($\sim 2 \times 10^{12}$ cm$^{-2}$) for such compensated samples is observed and can contribute to low RF loss. An exceptionally high effective resistivity of above 8 kΩ.cm is achieved, corresponding to an RF loss below 0.3 dB/mm at 10 GHz.

**Main text.**

GaN high electron mobility transistors (HEMTs) integrated on Si substrates promise to enable attractive RF performance of GaN technology at low-cost [1]. However, compared with a semi-insulating substrate such as SiC or Sapphire, the use of Si can lead to increased substrate-induced RF losses and harmonic distortion (HD) even when high-resistivity (HR) Si is used [2]. A lossy, non-linear substrate can also negatively impact the performance of active RF circuits such as RF switches and power amplifiers [3], [4]. The root cause for substrate loss in GaN-on-Si technology has been widely studied in recent years and is now accepted to be from the presence of a conductive channel in Si [5], [6], [7]. During III-N growth on Si, a parasitic surface conduction (PSC) layer forms close to the Si surface, decreasing the effective substrate resistivity ($\rho_{eff}$) to values that can be orders of magnitude lower than the nominal HR Si resistivity. More precisely, Al and, in lesser amount, Ga atoms diffuse into Si, resulting in a p-type doping of the top Si layer [8]. Other factors, such as spontaneous or piezoelectric polarization or impurity activation (interstitial O, N, H etc.) can also contribute independently of the diffusion of dopants [9]. This PSC layer must be suppressed to reduce the substrate loss. The most critical step is known to be the growth of the AlN nucleation layer [7]. Indeed, most of the parasitic doping takes place during this step, and the electrical properties of the AlN/Si interface (charge, defects) strongly influence the PSC layer conductivity. Controlling the AlN nucleation step is crucial to enable highly resistive GaN-on-Si stacks and approach the semi-insulating properties of GaN-on-SiC.
Mitigation of RF loss at the AlN/Si interface is a widely studied topic [5], [7], [8], [9], [10], [11], [12], [13], [14], [15]. However, when parasitic doping is kept under control, the PSC layer polarity seems to be dependent on the MOCVD conditions, with some groups reporting an n-type layer [7], [9], [13] and others a p-type layer [5], [11], [12]. Consequently, proposed solutions might only apply in a given process window and in a particular reactor. For instance, the impact of an interfacial silicon nitride ($SiN_x$) is seen as beneficial in [11] but detrimental in [7]. Some mitigation techniques involve additional process steps such as compensation doping, which is impractical from doping control perspective [10], [15]. More recently, the in-situ pre-treatment of Si surface by trimethylaluminum (TMAl) has been found to strongly influence the RF performance [14], [16], [17], although those works have not identified the underlying electrical reason for RF loss fluctuation. In this letter, we propose a more complete physical understanding of the AlN/Si interface. The impact of TMAl predose on the interfacial layer composition is linked to the electrical properties of the interface and to the RF loss performance. The understanding is validated by modelling and fabrication of an ultra-low-loss AlN/Si stack. A method involving no additional process step and limited characterization and experiment is finally proposed to consistently obtain good RF performance.

Two series of 200 mm AlN/Si samples produced in different cleanrooms and MOCVD reactors (Reactors A and B) are studied. All samples consist of a 175 nm-thick AlN layer grown in separate AIXTRON G5+ C Planetary reactors at 1010 °C, which was previously optimized to limit Al diffusion while maintaining good crystal quality [5]. The starting Si wafers are 200 mm high-resistivity Si (> 4 kΩ·cm) with interstitial oxygen concentration ($[O_i]$) lower than 5 p.p.m.a (specified by the supplier) for Reactor A and unspecified $[O_i]$ for Reactor B. Exact resistivities are not provided by the supplier and are known to change after MOCVD due to activation of $O_i$-related thermal donors [2]. In both series, the TMAl predose was varied from 7.7 $cm^3$ to 61.3 $cm^3$ as described in Table I. Material characterization was performed using X-ray Diffraction (XRD) to ensure the good crystalline quality of the nucleation layers fabricated in Reactor A. Full widths at half maximum (FWHM) of 1472 arcsec, 1394 arcsec and 1582 arcsec are obtained for samples A8, A31 and A61, respectively, from omega scans of the AlN (002) diffraction peak. Root mean square (RMS) roughness, measured by atomic force microscopy (AFM), is contained to less than 0.22 nm for all samples. The potential change in thermal resistance ($R_{th}$) of the AlN layers or AlN/Si interface was not studied here as their contribution to the $R_{th}$ of the total GaN-on-Si stack is expected to be small compared with the AlGaN buffer layers [18].

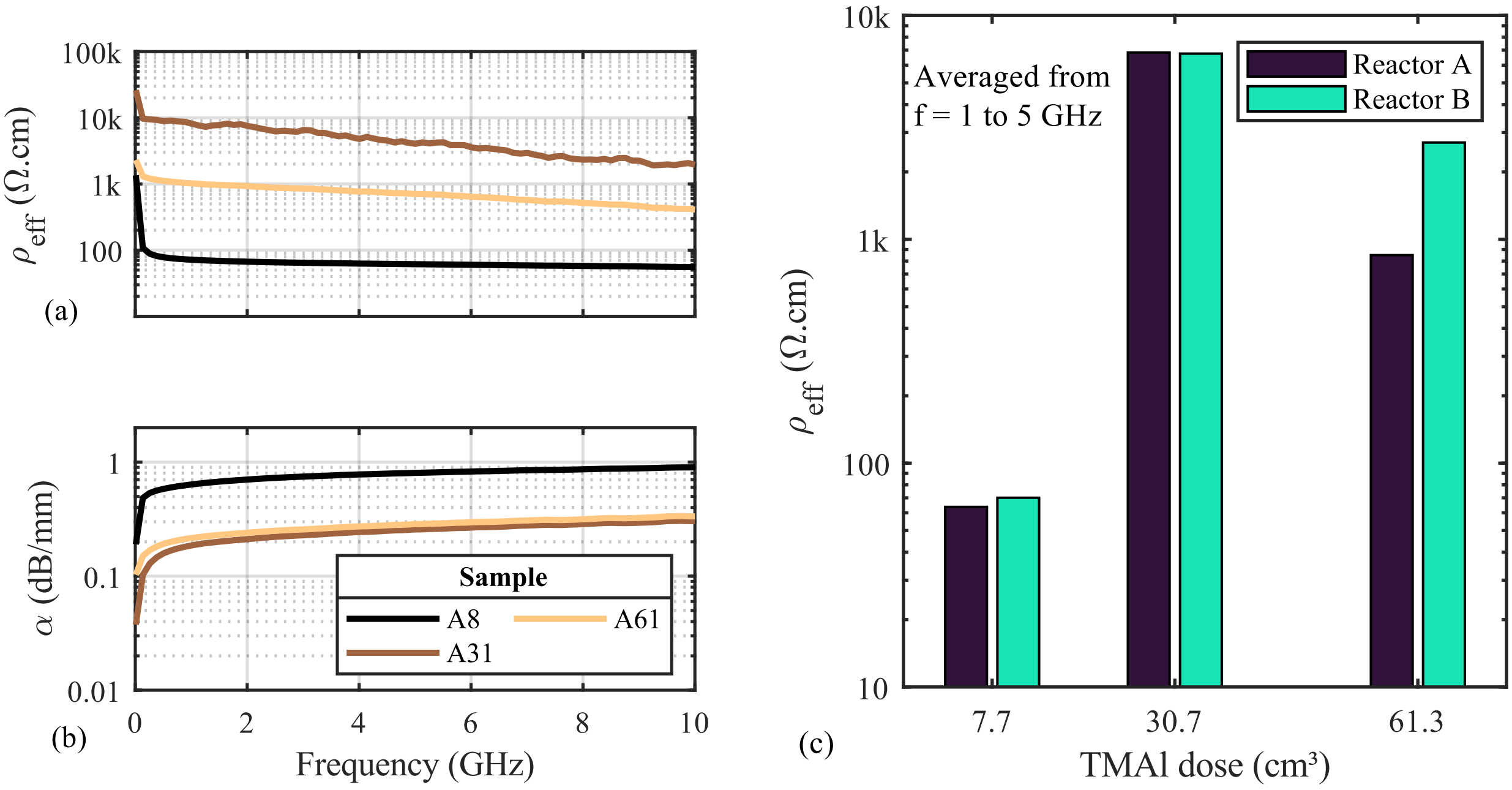

Fig. 1. (a) Effective resistivity and (b) total insertion loss for the CPW lines on representative dies on AlN/Si samples produced in Reactor A. (c) TMAl predose strongly impacts $\rho_{eff}$. The effect is independent of the reactor or starting Si wafer. W/S = 17 µm/10 µm and 36.6 µm/20 µm for Reactor A and B, respectively.

Following AlN growth, coplanar waveguide (CPW) lines were patterned directly on the AlN surface, also in separate labs and with different metal stacks. A signal line width (W) and signal-to-ground spacing (S) of 17 µm and 10 µm, respectively, are used for samples of Reactor A. For Reactor B, W = 36.6 µm and S = 20 µm. S-parameters measurements were performed in dark conditions at 25 °C and the effective substrate resistivity was extracted following the procedure described elsewhere [19]. The effective resistivity figure-of-merit is given by $\rho_{eff} = \frac{F_{bot}}{G}$, in which $G$ is the shunt conductance of the transmission line and $F_{bot}$ is a geometrical factor that accounts for the field repartition between the top and the bottom of the CPW line. For such $S \gg t_{AlN}$, the dimension-induced difference in $\rho_{eff}$ is expected to be limited [20].

Table 1. Description of the different samples used in this study.

| TMAl predose (cm³) | Sample denomination | |
|---|---|---|
| | Reactor A | Reactor B |
| 7.7 | A8 | B8 |
| 30.7 | A31 | B31 |
| 61.3 | A61 | B61 |

The results are shown in Fig. 1a. $\rho_{eff}$ is extracted below 5 GHz, relevant for Frequency Range 3 (FR3) applications [21]. At higher frequencies, the extraction suffers from metrology-related effects discussed in [19]. A strongly non-linear impact of TMAl predose on substrate RF performance is observed.An improvement by ~2 orders of magnitude in $\rho_{eff}$ is achieved from sample A8 ($\rho_{eff} \cong 60\ \Omega\cdot\text{cm}$ ) to sample A31 ($\rho_{eff} \cong 7\ \text{k}\Omega\cdot\text{cm}$ ), and $\rho_{eff}$ then decreases when the predose is further increased to sample A61 ($\rho_{eff} \cong 900\ \Omega\cdot\text{cm}$ ). $\rho_{eff} > 5\ \text{k}\Omega\cdot\text{cm}$ is extracted for sample A31, corresponding to a quasi-lossless substrate [22]. The CPW losses (α), which include both the metallic and substrate losses and are depicted in Fig. 1b for the three samples. For Sample A31, the loss is contained below 0.31 dB/mm at 10 GHz and below 0.46 dB/mm at 40 GHz (not shown). The benefits of using the $\rho_{eff}$ figure-of-merit is clear: for samples A31 and A61 where series metallic losses represent most of the loss, α is practically independent of substrate loss but $\rho_{eff}$ allows to discriminate a significant difference in substrate performance, while being independent of the CPW metallization scheme. Such a difference in $\rho_{eff}$ can cause a ~30 dB difference in substrate HD [3], [23]. This effect can be accurately reproduced in a different reactor and for different starting HR Si substrates (Fig. 1c). In the following, additional physical characterization will help to understand the origin of this non-linear effect and to identify the TMAl predose for which $\rho_{eff}$ is the highest.

The significant difference in $\rho_{eff}$ indicates a change in Si surface conductivity, which could be caused by doping. Al in-diffusion and consequent Si doping is known to be affected by process conditions [5]. To measure the surface doping, Spreading Resistance Profiling (SRP) measurements were taken on all three samples from Reactor A. SRP data can reveal the presence of ionized dopants resulting from Al diffusion during MOCVD growth for depths into Si ranging from ~20 nm to a few µm. However, fixed charges or traps located at the interface or in the AlN layer cannot be revealed by SRP as the AlN layer is removed for the measurement of the bevelled Si surface. The spreading resistance profiles in Fig. 2 show a limited Al in-diffusion for the three samples and the resistivity stays above ~60 Ω·cm for the entire profile. The dopant profile is following Fick's second law for all samples, with diffusion lengths of 43 nm, 50 nm and 88 nm for samples A8, A31 and A61, respectively. A slightly increased dopant diffusion is seen for sample A61. The charge corresponding to the increased surface doping can be calculated as:

$$Q_{dop} = \int_0^{t_{PSC}} (N_A - N_{A,bulk})dy,$$

in which $N_{A,bulk}$ is the doping far from the surface and $t_{PSC}$ is the PSC layer thickness, defined as the depth at which $N_A = N_{A,bulk}$. Integrated doping results are shown in the inset of Fig. 2. The difference in charge between the three samples is lower than $5 \times 10^9 \text{cm}^{-2}$ and is not sufficient to explain any difference in $\rho_{eff}$, as could be confirmed by TCAD simulations. Also, all samples show p-type conductivity for the entire profile (i.e. until ~20 µm depth), meaning that no significant interstitial oxygen-related thermal doping (which would be n-type) activation is taking place.

Furthermore, STEM and EDS images indicate the presence of a $SiC_xN_y$ layer between AlN and Si [16]. As seen in Fig. **3**, the composition of this ~5 nm-thick layer measured with EDS changes from a $SiN_x$-like layer to a $SiC_x$-like layer when TMAl predose is increased. Carbonization of the Si surface has been previously linked to TMAl predose in [24]. Such an interfacial layer cannot be detected by SRP and additional electrical characterization is required to understand its electrical impact.

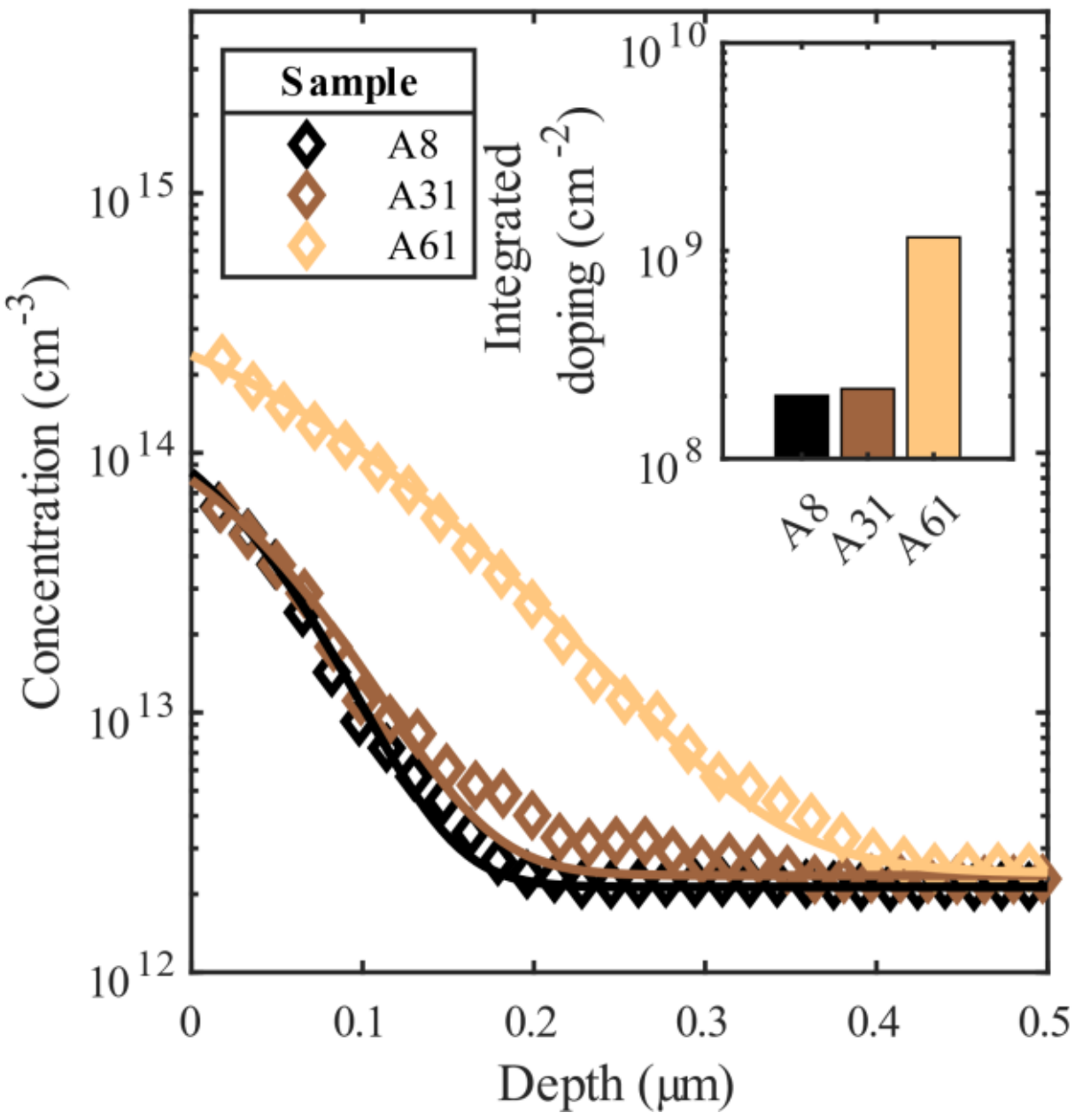

Fig. 2. SRP measurement for all three AlN/Si samples of Reactor A. Best fit for Fick's second law is included in continuous lines. Inset: the sheet carrier concentration induced by the surface doping is lower than $2 \times 10^9 \text{cm}^{-2}$ for the three samples. The SRP data shows p-type conductivity for the full profile in all samples.

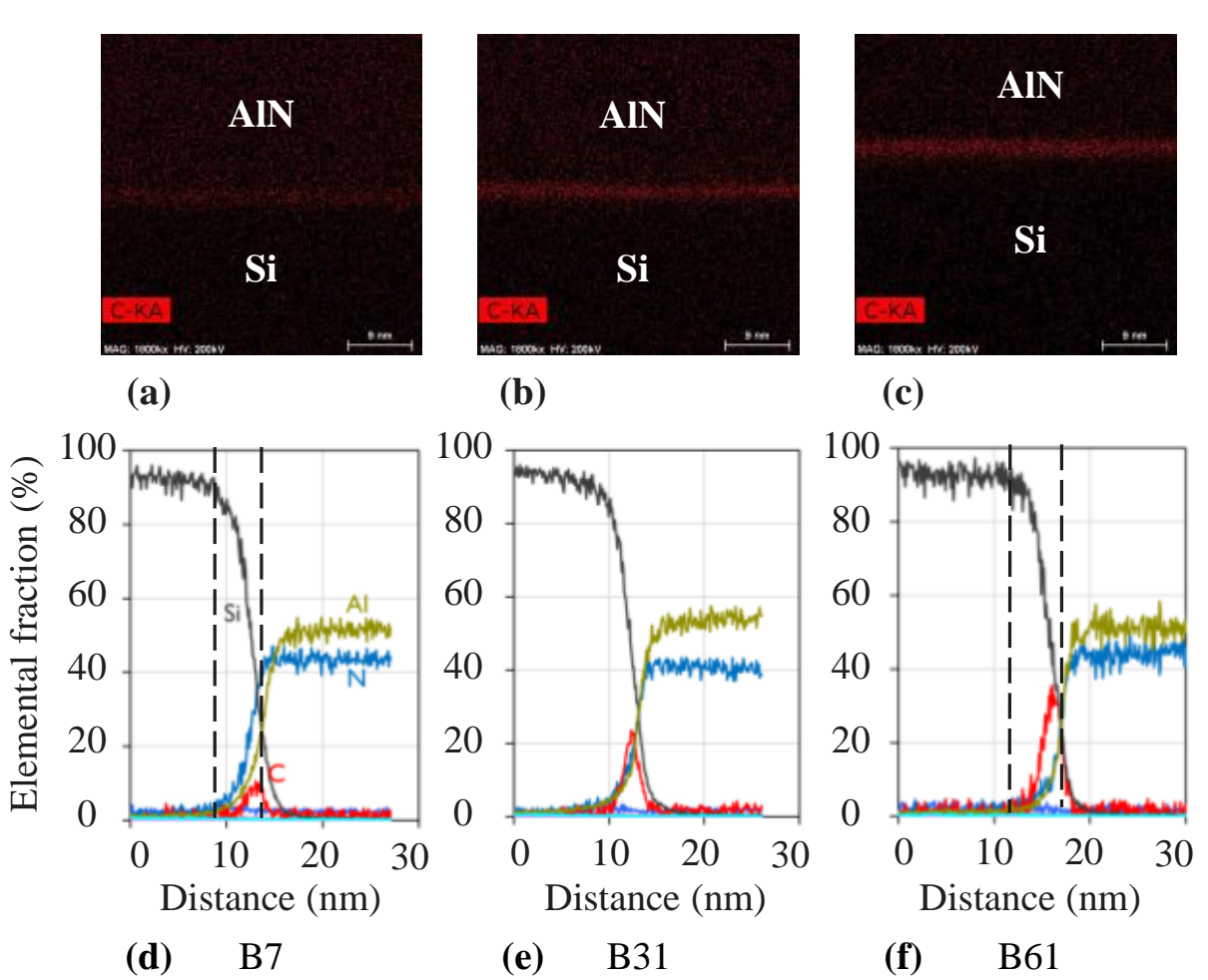

Fig. 3. Physical characterization of Reactor B samples. (a-c) STEM images reveal the presence of an interfacial layer between AlN and Si. (d-f) EDS scan along a cross-section of the AlN/Si interface show an evolution of interfacial layers' composition as TMAl predose is increased. Lower predose favours the formation of SiN, while higher predose produces a higher fraction of SiC at the interface.

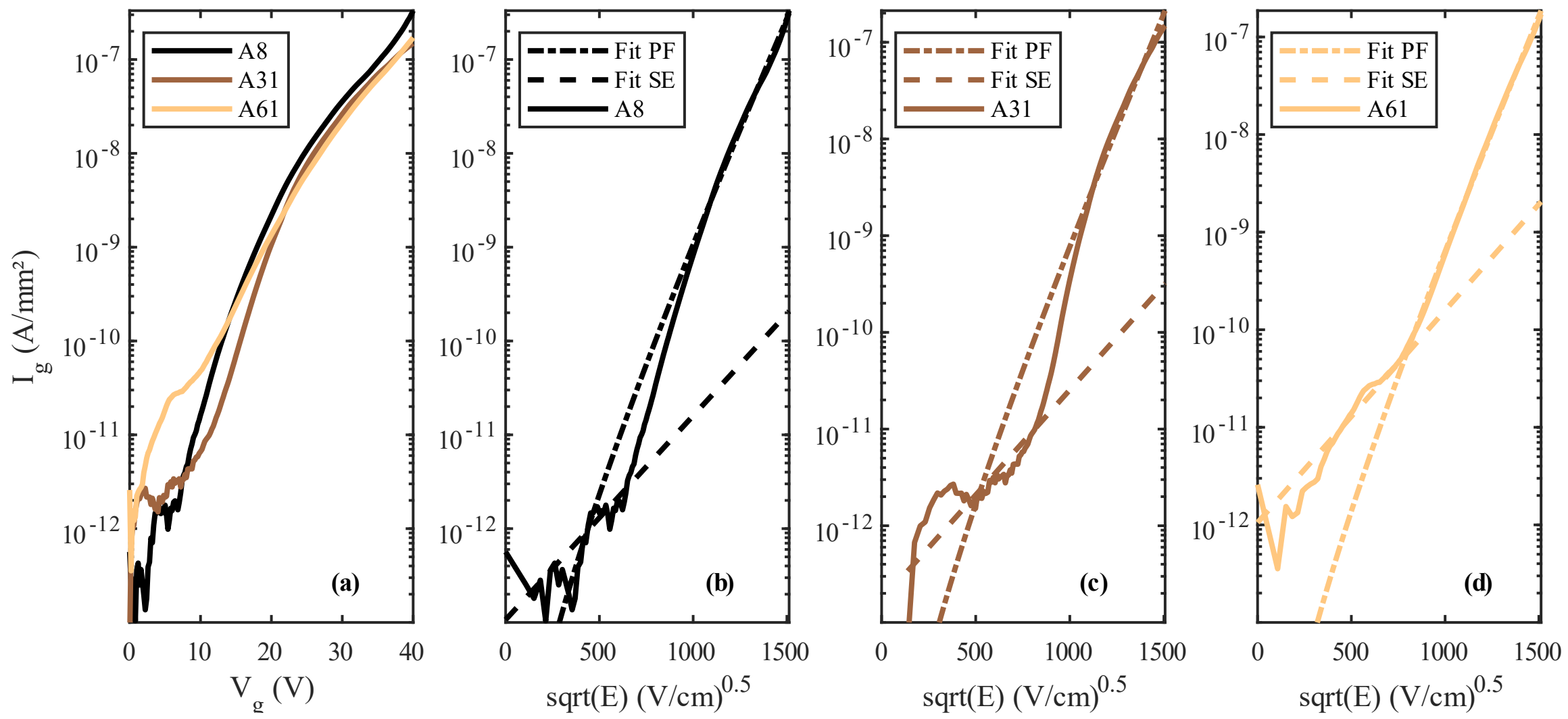

Fig. 4. (a) I-V sweeps for all samples from reactor A. (b-d) Leakage current plotted against sqrt(E) including best fits for SE and PF mechanisms for samples (b) A8, (c) A31, and (d) A61.

I-V sweeps were measured to get more insight on the vertical carrier transport mechanisms close to the interface and across the AlN layer. Results are shown in Fig. 4a and in Fig. 4b-d as a function of $\sqrt{\mathrm{E}}$, where the electric field E across the AlN layer was calculated as:

$$\mathrm{E}_{AlN} = \frac{V_{gate} - \phi_s}{t_{AlN}},$$

for which the surface potential $\phi_s$ was extracted from the C-V curves (see later). At low fields, the fit to Schottky Emission (SE) mechanism is made difficult by the high measurement noise at low current levels. The SE current is given by [25]:

$$J_{SE} \propto \exp\left[\frac{-q\left(\phi_B - \sqrt{\frac{qE_{AlN}}{4\pi\varepsilon_{AlN}}}\right)}{kT}\right],$$

in which $\phi_B$ is the barrier height and $\varepsilon_{AlN}$ the permittivity. Here $\varepsilon_{AlN} = 10$ was extracted from the C-V. Barriers heights close to 2.1 eV are extracted for all three samples (with a relatively high uncertainty linked to the noise). For E > ~1 MV/cm, a Poole-Frenkel (PF) mechanism can be well fitted to the experimental curves for all three samples. This behaviour is consistent with recent literature on MOCVD AlN/Si [13], [26]. In this bias region, the current is given by [25]:

$$J_{PF} = q\mu_{AlN}N_c E_{AlN}\exp\left[\frac{-q\left(\phi_T - \sqrt{\frac{qE_{AlN}}{\pi\varepsilon_{AlN}}}\right)}{kT}\right],$$

in which $\mu_{AlN}$ is the mobility inside AlN layer, $N_C$ the density of states in the conduction band and $\phi_T$ the depth of traps in AlN. $N_C$ was taken as $6.3 \times 10^{18}$ cm$^{-3}$ from [27]. For $\mu_{AlN}$, a value of 20 cm$^{-2}$V·s was used, which is consistent with heteroepitaxial AlN films with a dislocation density of $10^{9}$-$10^{10}$ cm$^{-2}$ [28]. The extracted activation energies are 1.08 eV, 1.09 eV and 1.10 eV for samples A8, A31 and A61, respectively, indicating the presence of a defect level in the AlN layer independently of the TMAl predose. While this defect energy is larger than what has been reported for AlN [29], it should be noted that the value of $\phi_T$ is sensitive to the $\mu_{AlN}$ used. $\mu_{AlN}$ is not expected to change with TMAl predose as the crystalline quality of the different layers is comparable. With no significant difference in activation energies, barriers or current densities, it is assumed in the following that the band alignment between AlN and Si is not modified by the TMAl predose and that the electrical properties of the AlN layer are also unaffected. We now investigate the possibility of a different interface fixed or trapped charge which could affect the conductivity of the Si surface.

Top-down metal-insulator-semiconductor (MIS) capacitor C-V measurements were taken at a frequency of 1 kHz on dot capacitors (area: 0.4 mm²) and are shown in Fig. 5a. A significant flatband voltage ($V_{FB}$) shift difference is observed as the TMAl predose is modified. As SRP indicates negligible variation in surface doping and AlN electrical properties appear to be identical for all three samples, the entirety of the $V_{FB}$ shift can be attributed to interface charge. Interface fixed charge can be extracted as:

$$\Delta Q_{int} = C_{AlN}V_{FB},$$

in which $C_{AlN}$ is the accumulation capacitance and $V_{FB}$ is estimated from the max $\left(\frac{d^2C^{-2}}{dV^2}\right)$ criterion [30]. For the samples which present hysteresis, an average between forward and reverse sweep is used. Interestingly, $Q_{int}$ presents a linear evolution with TMAl predose, as seen in Fig. 5b where charges extracted from samples of Reactor B are also included. For low predose (sample A8), a native positive charge exists in the AlN/Si system. This charge could either be natively present in the $SiN_x$ layer or could consist of AlN spontaneous polarization charge, which is positive and of the order of ~$10^{13}$ $cm^{-2}$ [9], [13]. Then, as predose is increased, the gradual increase of the C fraction in the interfacial layer leads to formation of negative interface charge compensating the initial $Q_{int}$. Further predose increase can make the interface strongly negatively charged, as in sample A61. A large interface charge (positive or negative) attracts free carriers (electrons or holes, respectively) to the Si surface and causes a decreased $\rho_{eff}$. For sample A31, the lowest remaining interface charge leads to the highest $\rho_{eff}$.

A positive interface charge for low predose also explains the anomalous "low-frequency" C-V response in inversion regime of sample A8. Indeed, the minority carrier response time in silicon is typically larger than 1 ms and thus, at an AC frequency of 1 kHz only majority carriers at the depletion layer edge can respond, leading to a low measured capacitance in inversion region. However, with highly positive $Q_{int}$, an electron inversion layer is present across the entire wafer and in particular away from the gate electrode. When the capacitor is biased in inversion, a conductive layer larger than the gate electrode area is accessible for the AC current to flow and provides a source of minority carriers. The capacitance associated with the extended conductive layer is large and increases the value of the depletion capacitance. Consequently, the measured capacitance rises to $C_{AlN}$.

For accurate TCAD modelling, the interface trap density ($D_{it}$) was extracted for the curves in Fig. 5a using the Terman method [31]. Ideal C-V curves were constructed in TCAD [32] using the SRP doping profiles of Fig. 2 to compute the "trap-free" $\phi_s - V_g$ relationship. For the experimental curves, $\phi_s$ is extracted following the method described in [33].

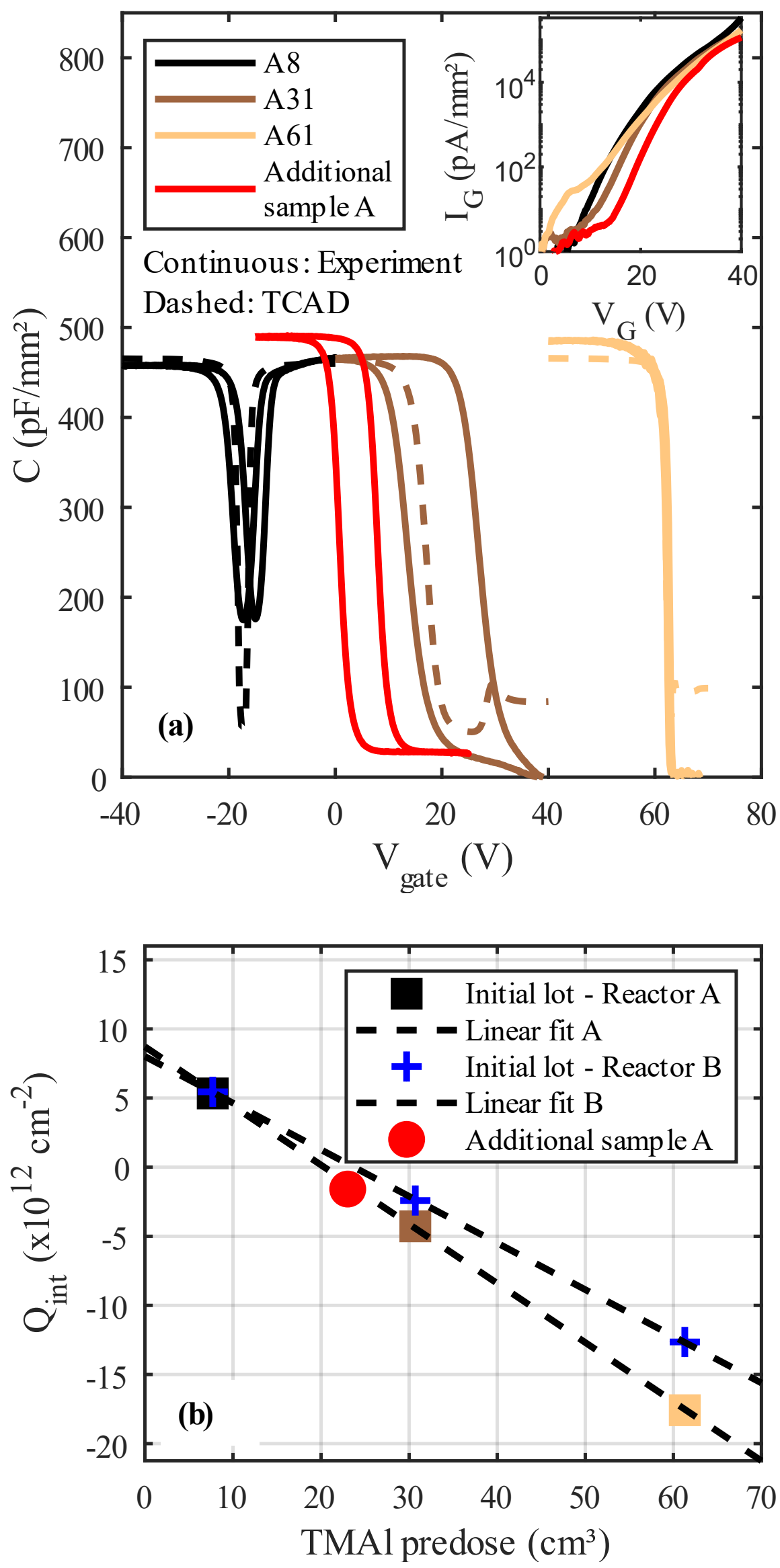

Fig. 5. (a) Top-down C-V measurements of the AlN/Si stacks for samples of Reactor A. Vastly different trends are observed depending on the TMAl predose. TCAD modelling considering trap charges, SRP doping and interface traps shows a good match to experiments. f=1 kHz. Inset: corresponding vertical I-V curves. (b) Interface fixed charge extracted from the C-V curves of all samples plotted against TMAl predose. A linear dependence is revealed. The additional sample with intermediate predose confirms the linear trend.

The difference between ideal and measured $\phi_s - V_g$ then provides a rough estimate of $D_{it}$. $D_{it}$ of $\sim 6 \times 10^{11}$ cm$^{-2}$eV$^{-1}$, $\sim 2 \times 10^{12}$ cm$^{-2}$eV$^{-1}$ and $\sim 4 \times 10^{11}$ cm$^{-2}$eV$^{-1}$ are extracted at midgap for samples A8, A31 and A61, respectively. When those traps are added as bulk defects into the 5 nm-thick interfacial layer, a reasonably good match between the TCAD model and experiments is obtained. The higher $D_{it}$ for sample A31 can further contribute to the high $\rho_{eff}$. As the Fermi level at the Si surface is pinned close to midgap by the high density of traps, the formation of a conductive layer is prevented, and the effective resistivity is high.

The physical arguments presented above are finally confirmed by fabrication of an additional sample in Reactor A. A predose of ~23 cm³ was chosen to approach the ideal case of no interface charge in Fig. 5b. All other process conditions are kept unchanged. As shown in Fig. 5b, the interface charge extracted from C-V measurements of this additional sample follows the linear relationship found for the initial lot. Furthermore, RF measurements reveal exceptionally high $\rho_{eff}$ > 8 kΩ.cm which is above the value of A31 (Fig. 6a and c). The material characterization reveals an interface composition which is in line with the other samples (Fig. 6b and d), and the I-V is also in line with other samples (see inset of Fig. 5b. Those learnings suggest a procedure to reach highly resistive AlN/Si stacks without needing a complex design of experiment (DOE) or the use of additional process steps.

(i) Production of one sample with low TMAl predose, corresponding to positive charge in the interfacial layer;
(ii) Production of one sample with large predose, corresponding to highly negative charge;
(iii) Extraction of the interface charge for samples (i) and (ii);
(iv) Linear interpolation to determine the precise predose leading to the close-to-ideal case of no net interface charge

Furthermore, we expect that the methodology described here can be applied for complete HEMT stacks as the upper III-N layers only slightly shift the y-intercept of Fig. 5b by formation of additional predose-independent charge. Indeed, it has been observed that the RF loss of full HEMT stacks and the corresponding AlN/Si stacks can be correlated [8].

In conclusion, we proposed a complete physical and electrical description of the AlN/Si interface. For samples where doping at Si surface is kept under control by limiting Al diffusion, a significant PSC layer can still be formed by interface charges. While a certain amount of positive charge appears to be inherent to the AlN/Si system, additional negative charge is induced by the formation of a $SiC_xN_y$ interfacial layer between AlN and Si. The amount of negative charge was observed to be linearly dependent on the TMAl predose and can be linked to the evolution of the C concentration in this layer. Those learnings point to a simple procedure for reliable fabrication of low-loss GaN-on-Si stacks.

**Data availability statement**

The data that support the findings of this study are available from the corresponding author upon reasonable request.

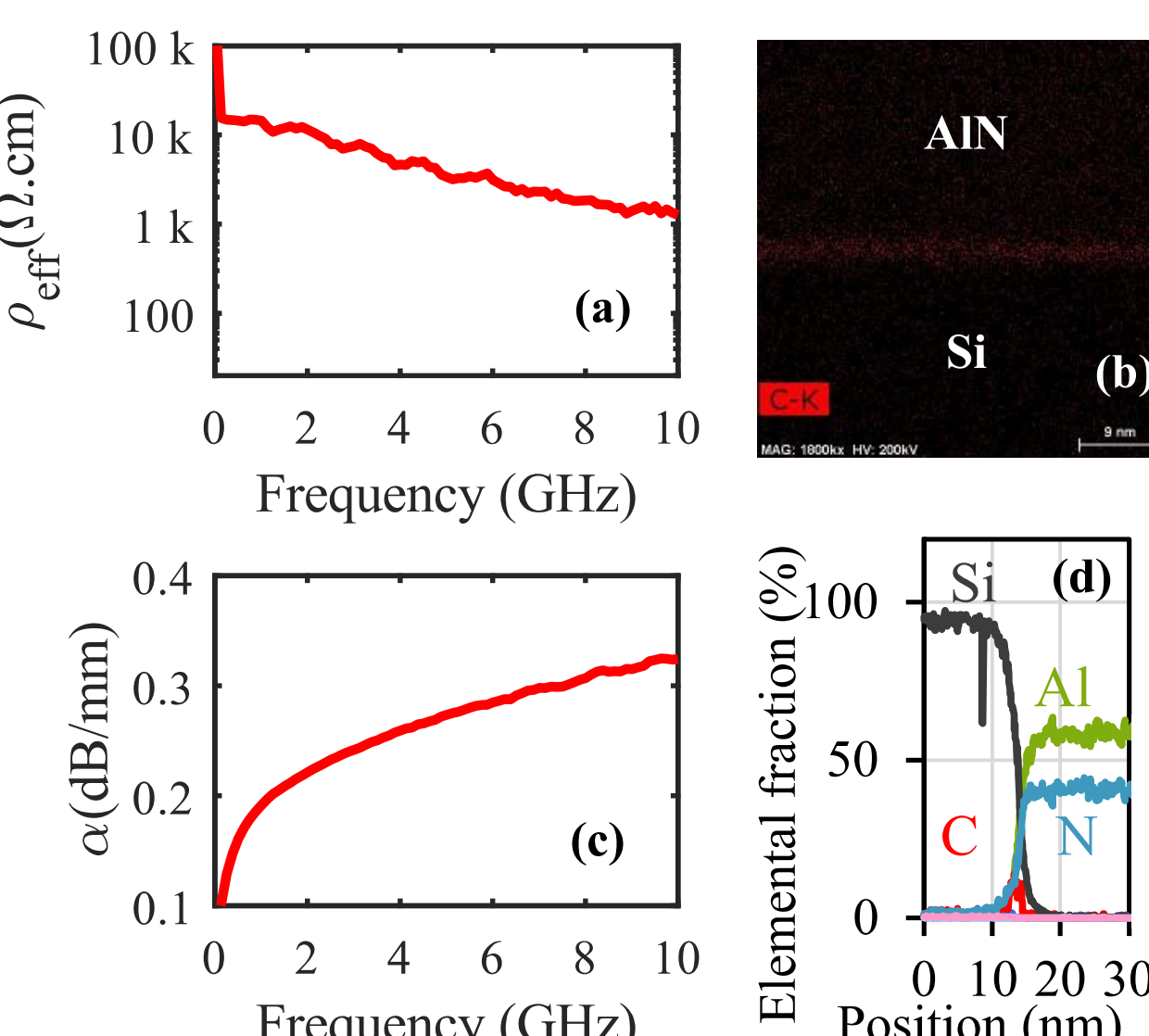

Fig. 6. RF and EDS characterization of the additional sample fabricated in Reactor A. (a) Effective resistivity (W/S = 17/10). (b) Interfacial layer identification. (c) Insertion loss. (d) The C fraction at the interface is intermediate between Samples A8 and A31.